\documentclass{article}
\usepackage{arxiv}  % Add 'final' option
\usepackage{cite}
\usepackage{graphicx}
\usepackage{amsmath,amssymb}
\usepackage{url}
\usepackage{booktabs}
\usepackage{algorithm}
\usepackage{algpseudocode}

\title{Recursive Language Models for Jailbreak Detection: A Procedural Defense for Tool-Augmented Agents}
\author{
  Doron Shavit  \\
  Silverfort.com \\
  \texttt{doron.shavit@silverfort.com}\\
}

\begin{document}
\maketitle

\begin{abstract}
Jailbreak prompts are a practical and evolving threat to large language models (LLMs), particularly in agentic systems that execute tools over untrusted content. Many attacks exploit long-context hiding, semantic camouflage, and lightweight obfuscations that can evade single-pass guardrails. We present RLM-JB, an end-to-end jailbreak detection framework built on \emph{Recursive Language Models} (RLMs)\cite{ref16}\cite{ref23}, in which a root model orchestrates a bounded analysis program that transforms the input, queries worker models over covered segments, and aggregates evidence into an auditable decision. RLM-JB treats detection as a procedure rather than a one-shot classification: it normalizes and de-obfuscates suspicious inputs, chunks text to reduce context dilution and guarantee coverage, performs parallel chunk screening, and composes cross-chunk signals to recover split-payload attacks. On AutoDAN-style adversarial inputs, RLM-JB achieves high detection effectiveness across three LLM backends (ASR/Recall 92.5--98.0\%) while maintaining very high precision (98.99--100\%) and low false positive rates (0.0--2.0\%), highlighting a practical sensitivity--specificity trade-off as the screening backend changes.
\end{abstract}


\section{Introduction}
Reproducible evaluation has become central to jailbreak research, enabling systematic comparison of attacks and defenses across models and threat models \cite{ref24,ref23}. Standardized benchmarks such as HarmBench combine curated harmful query sets, attack implementations, and refusal-robustness metrics to make red-teaming more comparable across systems \cite{ref13}. JailbreakBench complements this direction with an open robustness benchmark and leaderboards for consistent head-to-head comparisons \cite{ref1}. At the same time, real-world jailbreak communities and automated optimization procedures continue to produce adaptive attacks that evade static filters and single-pass detectors \cite{ref14,ref5,ref6}.

This work proposes \textbf{Recursive Language Models (RLMs)} as a security engineering pattern for jailbreak detection and introduces \textbf{RLM-JB}, a procedural detector designed for tool-augmented, stateful environments. The architectural premise is to externalize large inputs into a computation environment and have a root model orchestrate analysis via code execution, targeted submodel calls, and iterative state updates. This design is well matched to jailbreak defense because many modern attacks exploit context hiding, semantic camouflage, and split payloads that are difficult to capture reliably in a single global pass.

\textbf{Contributions.} (i) We describe an RLM-based detection architecture that turns jailbreak detection into a bounded analysis program with explicit intermediate artifacts; (ii) we present a practical RLM-JB pipeline that combines de-obfuscation, coverage-guaranteeing chunking, parallel screening, and compositional evidence aggregation; and (iii) we evaluate RLM-JB across multiple screening backends and report deployment-relevant metrics (ASR/Recall, FPR, Precision, F1).

\section{Related Work}
\subsection{Jailbreak attacks and automated optimization}
Empirical studies document recurring jailbreak strategies, including direct instruction hijacking, role-play and persona escalation, encoding/obfuscation, and long-context hiding \cite{ref5,ref15,ref17}. Automated optimization has become a particularly important driver of attack success: universal or transferable adversarial suffixes can transfer across models and tasks \cite{ref24}, while nested scenario-wrapping prompts generalize attacks and systematically evade safety filters \cite{ref5}. AutoDAN introduced a genetic-search approach to generate stealthy jailbreak prompts \cite{ref14}; later work extends this direction to multi-turn and more adaptive settings \cite{ref9}. These results motivate defenses that reason procedurally over structure, transformations, and multi-part payloads rather than relying only on superficial features.

\subsection{Prompt injection and defenses}
Prompt injection defenses span structured query reformulations \cite{ref2}, preference-optimization approaches \cite{ref3}, and architectural ``defense by design'' methods that aim to reduce injection surfaces \cite{ref4}. Detection-focused approaches include attention-based tracking of suspicious regions \cite{ref7}, unified defenses that target multiple adversarial classes \cite{ref10}, and practical prompt-injection detection frameworks \cite{ref8}. These systems vary in assumptions and deployment costs; RLM-JB is complementary in that it emphasizes coverage, normalization, and compositional evidence aggregation under a root-orchestrated procedure.

\subsection{Benchmarking and evaluation metrics}
Beyond HarmBench \cite{ref13} and JailbreakBench \cite{ref1}, recent evaluations emphasize the operational trade-off between blocking adversarial inputs and avoiding over-blocking benign traffic \cite{ref18,ref21}. Accordingly, we report both adversarial success metrics (ASR/Recall) and standard classification metrics (FPR, Precision, F1) to quantify robustness and usability under mixed benign/adversarial workloads.

For completeness, we also draw on a broader set of recent work on jailbreak prompting, automated red-teaming, prompt injection, and guardrail evaluation that informs the benchmark design and threat model considered here \cite{ref1,ref2,ref3,ref4,ref5,ref6,ref7,ref8,ref9,ref10,ref11,ref12,ref13,ref14,ref15},\cite{ref16,ref17,ref18,ref19,ref20,ref21,ref22}.

\section{Flow Overview}
\begin{figure}[t]
\centering
\includegraphics[width=0.68\linewidth]{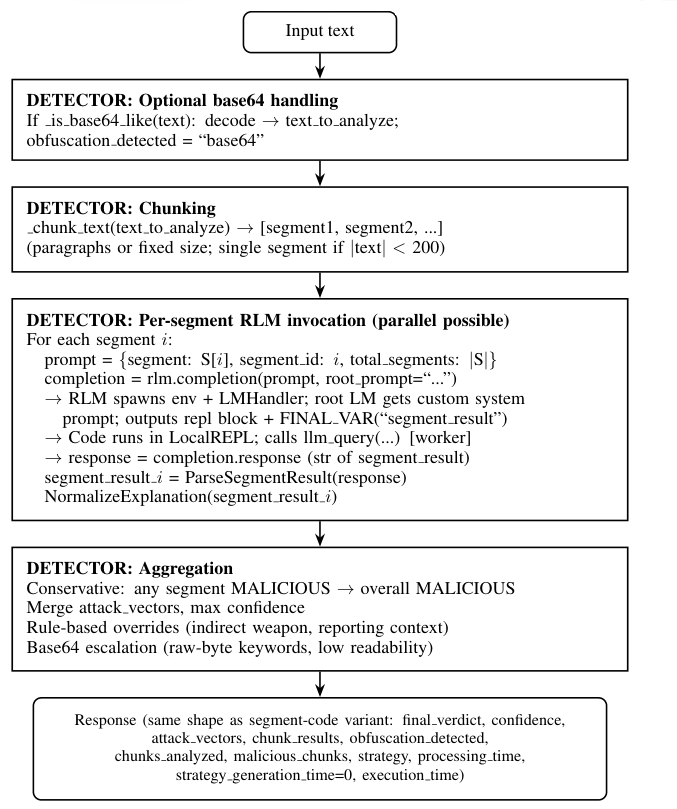}
\caption{RLM-JB flow chart.}
\label{fig:rlmjb-flow}
\end{figure}

Fig.~\ref{fig:rlmjb-flow} summarizes the RLM-JB pipeline as a staged procedure that transforms untrusted text into an auditable decision. The root model begins by ingesting the raw input (user prompt, retrieved context, or tool output) and applying lightweight canonicalization (e.g., normalizing whitespace and separators) so that superficial formatting changes do not affect downstream heuristics. It then performs targeted de-obfuscation checks: when the input contains high-signal encoding patterns (such as Base64-like spans), the system attempts decoding and carries forward both the original and decoded representations for analysis.

Next, the pipeline enforces \emph{coverage} by chunking the (possibly decoded) text into overlapping segments. Each segment is screened independently by worker calls that return a structured assessment (malicious/benign), confidence, and extracted signals (e.g., instruction hijacking cues, persona escalation prompts, or tool-manipulation attempts). Segment-level screening reduces susceptibility to long-context hiding and ``lost in the middle'' effects, and it yields localized evidence that can be inspected and audited.

Finally, the root model aggregates segment-level evidence to recover split-payload structures and reach a conservative global decision. If multiple segments contribute complementary indicators (for example, one segment supplies an ``ignore instructions'' preamble while another contains the actionable payload), the aggregator links them into a single attack hypothesis and produces a final verdict with a concise explanation. The flow also supports escalation and override logic, such as re-screening decoded text when encoding evidence is strong, and it returns a structured output that includes the verdict, confidence, and supporting evidence for downstream policy enforcement.

\section{Threat Model and Problem Formulation}
We consider a tool-augmented agent that performs user tasks in stateful environments (e.g., e-mail, collaboration channels, and transactional workflows). The attacker can inject or control untrusted text that the agent will process (e.g., retrieved messages, documents, or third-party content) and seeks to override policies or induce unsafe tool behavior. The defender places a jailbreak detector in front of the agent to label an input as \emph{malicious} (block or sanitize) or \emph{safe} (allow). We evaluate detection as binary classification with standard confusion-matrix metrics and treat an adversarial input as successful if it causes a refusal failure or policy-violating behavior in the evaluation harness.

\section{RLM-JB: Recursive, Procedural Jailbreak Detection}
\subsection{Recursive Language Models}
An RLM \cite{ref16} couples an LLM with a constrained sandbox environment for safe, bounded execution. A \emph{root} model orchestrates analysis by emitting code, calling worker models, and updating state iteratively, producing a final structured verdict. This structure supports recursion over parts of the input (segment-by-segment analysis), deterministic transformations (normalization/de-obfuscation), and auditable intermediate artifacts (per-segment verdicts and extracted attack vectors).

\subsection{Detection pipeline}
RLM-JB treats jailbreak detection as a procedure rather than a one-shot classification. First, it normalizes the input and applies de-obfuscation when encoding patterns are detected (e.g., Base64-like strings). Second, it chunks long inputs to guarantee coverage and reduce context dilution; this step is designed to counter long-context hiding and ``lost in the middle'' effects. Third, it screens each chunk with a worker model (parallelizable) that returns a chunk verdict, confidence, and extracted attack vectors. Finally, it aggregates evidence conservatively across chunks to recover split payloads, producing a final verdict and explanation.

\section{Experimental Settings}
We evaluate RLM-JB using AutoDAN-style benchmarks that simulate realistic, stateful environments and multi-step tool use, where agents may execute multiple tool calls for each task per task \cite{ref14}. We run 400 adversarial test cases per screening backend and interleave 200 benign prompts to estimate over-blocking. Some LLMs cannot be used as screening backends because they do not support code execution; for example, we evaluated models such as Llama-4-Maverick but exclude them from the quantitative comparison because the RLM-JB pipeline cannot run without code-execution support. We further tested RLM-JB on newer prompt-injection payloads collected from public prompt-injection catalogs (e.g., InjectPrompt) and on common permutations of each payload \cite{ref27}. 

\subsection{Metrics}
Let TP, FP, TN, FN denote true/false positives/negatives. We report ASR/Recall ($\frac{\text{TP}}{\text{TP}+\text{FN}}$) on adversarial cases, FPR ($\frac{\text{FP}}{\text{FP}+\text{TN}}$) on benign cases, Precision ($\frac{\text{TP}}{\text{TP}+\text{FP}}$), and F1. These metrics quantify both blocking performance and the operational cost of false alarms.

\section{Results}
Table~\ref{tab:models} summarizes performance across three screening backends. RLM-JB achieves high Recall (92.5--98.0\%) and very high Precision (98.99--100\%) across all models. The dominant cross-model difference is in FPR, which increases from 0.0\% on DeepSeek-V3.2 to 0.5\% on GPT-4o and 2.0\% on GPT-5.2, indicating a sensitivity--specificity trade-off as the screening model changes.

\begin{table}[t]
\caption{Model evaluation across different LLM backends.}
\label{tab:models}
\centering
\begin{tabular}{lccc}
\toprule
Metric & DeepSeek-V3.2 & GPT-4o & GPT-5.2 \\
\midrule
ASR (Recall) [\%] & 92.50 & 97.00 & 98.00 \\
FPR [\%] & 0.00 & 0.50 & 2.00 \\
Precision [\%] & 100.00 & 99.74 & 98.99 \\
F1 Score [\%] & 96.10 & 98.35 & 98.49 \\
\bottomrule
\end{tabular}
\end{table}

\begin{table}[t]
\caption{GPT-5.2 baseline versus RLM-JB under AutoDAN.}
\label{tab:baseline}
\centering
\begin{tabular}{lcc}
\toprule
Metric & GPT-5.2 Baseline & GPT-5.2 + RLM-JB \\
\midrule
ASR (Recall) [\%] & 59.57 & 98.00 \\
FPR [\%] & 1.67 & 2.00 \\
Precision [\%] & 100.00 & 98.99 \\
F1 Score [\%] & 69.71 & 98.49 \\
\bottomrule
\end{tabular}
\end{table}

\noindent\textit{Note:} The false positive rate (FPR) was estimated using benign prompts generated by an LLM rather than real-world user traffic; it should therefore be interpreted as an approximate indicator of over-blocking under synthetic benign inputs.
\textbf{Additional InjectPrompt evaluation}. We further tested
RLM-JB on newer prompt-injection payloads from the Inject-
Prompt corpus and common permutations of each payload. In
this evaluation, RLM-JB detected all attacks with 100\% accu-
racy and produced zero false positives, indicating robustness
to recent injection techniques and their common variants. [27]

\section{Discussion}
The results indicate that procedural, multi-stage analysis can deliver robust jailbreak detection under realistic prompt transformations and multi-turn, tool-augmented settings. Across screening backends, Recall/ASR increases from 92.5\% (DeepSeek-V3.2) to 97.0\% (GPT-4o) and 98.0\% (GPT-5.2), indicating that stronger screening models improve sensitivity to adversarial intent. This gain comes with a gradual increase in over-blocking risk, with FPR rising from 0.0\% to 0.5\% and 2.0\%, respectively. Precision remains near-saturated (100.00\% on DeepSeek-V3.2, 99.74\% on GPT-4o, and 98.99\% on GPT-5.2), so F1 improves monotonically (96.10\% \(\rightarrow\) 98.35\% \(\rightarrow\) 98.49\%). Operationally, this suggests a practical tuning axis: GPT-4o provides a strong balance point (high Recall with low FPR), while GPT-5.2 maximizes sensitivity at the cost of additional benign blocking.

\textbf{Comparison to baseline AutoDAN handling.}
A key question is whether the gains are primarily attributable to the screening model or to the RLM-JB procedure. Evidence from AutoDAN-based baselines suggests the procedure is the dominant factor: chunk-level coverage, normalization/de-obfuscation, and cross-chunk evidence aggregation reduce the effectiveness of long-context hiding and split-payload strategies that can evade single-pass screening. In a baseline setting where GPT-5.2 is used directly (without RLM-JB’s chunking, normalization, and aggregation), the baseline achieves ASR/Recall of 59.57\%, FPR of 1.67\%, Precision of 100.00\%, and F1 of 69.71\% (Table~\ref{tab:baseline}). Under the same AutoDAN setting, RLM-JB improves ASR/Recall to 98.00\% (an \textbf{absolute gain of 38.43 points} and a \textbf{+64.5\% relative} improvement), while keeping Precision near-perfect (98.99\%) and maintaining a similarly low benign error rate (FPR 2.00\%). The corresponding F1 increase is substantial: from 69.71\% to 98.49\% (\textbf{+28.78 points}, \textbf{+41.3\% relative}). In addition, the
InjectPrompt-based evaluation suggests the procedure gener-
alizes beyond AutoDAN-style transformations to more recent
in-the-wild injection payloads and their surface-form variants,
though broader coverage and standardized reporting would
strengthen this conclusion \cite{ref27}. Taken together, these results
support the hypothesis that robustness is largely determined by
procedural coverage and compositional reasoning rather than
relying on a single model pass.

\textbf{Comparison to other published AutoDAN-inclusive defenses.}
To contextualize these results against recent guardrail benchmarking that includes AutoDAN-derived attacks, Adversarial Prompt Evaluation \cite{ref26} reports combined-set detection performance for widely used guardrails, including Granite Guardian 3.0 (Accuracy 0.960, F1 0.821, Recall 0.921, Precision 0.741) and Llama-Guard 2 (Accuracy 0.955, F1 0.758, Recall 0.693, Precision 0.836). Relative to these reported F1 values, RLM-JB’s best configuration (F1 98.49\%) corresponds to an \textbf{absolute gain of 16.39 points} over Granite Guardian 3.0 (\textbf{+20.0\% relative}) and an \textbf{absolute gain of 22.69 points} over Llama-Guard 2 (\textbf{+29.9\% relative}). Because evaluation corpora and label definitions differ across works, these numbers should be interpreted as directional evidence that RLM-JB’s procedural decomposition can substantially improve detection quality under AutoDAN-style transformations, rather than as a strict apples-to-apples leaderboard claim.

JBShield \cite{ref25} reports average detection Accuracy 0.95 and F1 0.94 across five LLMs and nine jailbreak attack families (including AutoDAN), with an average FPR around 2\%, and a mitigation variant that reduces average jailbreak ASR from 61\% to 2\%. In our AutoDAN setting, RLM-JB achieves comparable benign error rates (FPR as low as 0--2\%) while achieving higher F1 (up to 98.49\%), corresponding to a \textbf{+4.49 point absolute} improvement over 0.94 (\textbf{+4.8\% relative}). Concept-based single-pass defenses and RLM-style procedural analysis are not mutually exclusive: a practical deployment could use a lightweight detector (e.g., concept activation or pattern-based triage) as an early filter and reserve full RLM-JB analysis for high-risk inputs to reduce latency while preserving Recall.

Recent external work further motivates system-level defenses. Knowledge-driven multi-turn jailbreaking demonstrates that multi-turn frameworks can still achieve non-trivial attack success against advanced targets \cite{ref9}. This suggests that base-model alignment improvements do not eliminate jailbreak risk and reinforces the need for structured procedural defenses in agentic settings where failures can induce unauthorized tool actions.

\section{Limitations and Future Work}
First, while our evaluation is grounded in a realistic benchmark, it remains limited relative to fully adaptive adversaries who optimize directly against the detector \cite{ref6}. Second, RLM-JB introduces latency and cost due to chunking, normalization, and multi-call screening, which can constrain deployment in strict interactive settings. In particular, end-to-end processing time can increase substantially: in our setup, RLM-JB with GPT-5 incurred up to ~3× the processing time compared to the baseline GPT-5.2 pipeline. Third, generalization can vary with chunking policies, thresholds, prompt templates, and the choice of screening model. Future work. A natural next step is evaluation under multi-turn and agentic threat models, including indirect prompt injection, tool-output manipulation, and planning-time coercion, consistent with the attack capabilities emphasized in recent jailbreak and prompt-injection work. Additional directions include (i) adaptive red-teaming where attacks are optimized against the defense, (ii) calibrated risk scoring to support policy-driven decision thresholds, and (iii) engineering optimizations that reduce computational overhead and processing time while maintaining recall and a low FPR.

\section{Conclusion}
RLM-JB demonstrates that jailbreak detection can be strengthened by treating security as a procedural, recursive analysis problem. By combining de-obfuscation, coverage-guaranteeing chunking, chunk-level screening, and compositional evidence aggregation, the detector achieves strong performance on AutoDAN-style adversarial inputs with low false positives, supporting the claim that analysis procedure is a key determinant of robustness.

Code availability: The GitHub repository for this work will be published shortly.

\newpage
\appendix
\section*{Appendix A - Mathematical Formulation}

\subsection{Notation and Definitions}

\paragraph{Input Space.}
\[
\mathcal{X} = \text{space of all possible text inputs}
\]
\[
x \in \mathcal{X}, \quad \text{where } x = (x_1, x_2, \ldots, x_n) \text{ and } x_i \text{ are characters}
\]
\[
|x| = n \text{ denotes the length of input } x
\]

\paragraph{Output Space.}
\[
\mathcal{Y} = \{\text{MALICIOUS}, \text{SAFE}, \text{ERROR}\}
\]
\[
y \in \mathcal{Y} \text{ is the final verdict}
\]

\paragraph{LLM Roles (no strategy tuple).}
\begin{itemize}
\item \textbf{Root LM:} In segment-code variant, generates Python code once (cached by text key); in RLM per-segment variant, invoked per segment via $\text{RLM}.\text{completion}(\text{ctx}_i)$ and produces a repl code block that calls the worker.
\item \textbf{Worker LM:} Classifies a single segment; invoked via $\text{llmquery}(p_i)$ from generated code or from within RLM execution.
\end{itemize}

\[
\mathcal{F} = \text{focus/attack-vector space (threat categories)}
\]
\[
\mathcal{V} = \{\text{MALICIOUS}, \text{SAFE}\} \quad \text{(chunk verdicts)}
\]

\paragraph{Time.}
\[
T(n) = \text{total processing time for input of length } n
\]
\[
T_{\text{code}} = \text{code generation time (segment-code: once per text or cache hit; RLM per-segment: 0)}
\]
\[
T_{\text{worker}} = \text{worker LLM analysis time per chunk}
\]
\[
k = \text{number of chunks}
\]

\subsection{Stage 0: De-obfuscation}
The detector applies de-obfuscation before chunking. Only Base64 is supported; no strategy parameter.

\paragraph{Detection.}
\[
\text{IsBase64Like}(x)
= \mathbb{1}\!\left\{|x_{\text{strip}}| \ge N\right\}
\cdot
\mathbb{1}\!\left\{\text{ratio}_{\text{b64}}(x) \ge \delta \right\}
\]
where $x_{\text{strip}}$ is $x$ with whitespace normalized and $\text{ratio}_{\text{b64}}(x)$ is the fraction of characters in the Base64 alphabet.

\paragraph{De-obfuscation Function.}
\[
x' = \mathcal{D}(x) =
\begin{cases}
\text{Base64}^{-1}(x) & \text{if } \text{IsBase64Like}(x)\ \land\ \text{DecodeOK}(x) \\
x & \text{otherwise}
\end{cases}
\]
Set $\mathit{obf}=\text{"base64"}$ if decoding was applied, else $\mathit{obf}=\text{"none"}$. All downstream steps use $x'$ as the text to analyze.
\[
T_{\text{deobf}} = O(n)
\]

\subsection{Stage 1: Chunking}
Chunking is fixed in the detector; no master-generated strategy. Single segment for short text; otherwise paragraphs or fixed-size.

\paragraph{Chunking Function.}
\[
C_1, C_2, \ldots, C_k = \mathcal{C}(x')
\]
with
\[
\mathcal{C}(x') =
\begin{cases}
x' & \text{if } |x'| < N \text{ or } x' \text{ empty} \\
\text{Split}_{\text{paragraphs}}(x'; \min\text{L}=M) & \text{if yields } \ge 2 \text{ segments} \\
\text{Split}_{\text{fixed}}(x', \text{L}_{\text{max}}) & \text{otherwise}
\end{cases}
\]
$\text{Split}_{\text{paragraphs}}$: split on double newline, keep segments with length $> L$.

$\text{Split}_{\text{fixed}}(x', L)$: partition $x'$ into contiguous segments of size $L$ (last segment may be shorter).

We have $\bigcup_{i=1}^{k} C_i = x'$ and $k = |\mathcal{C}(x')|$. If $\mathcal{C}(x') = \emptyset$, set $C_1=x'$, $k=1$.
\[
T_{\text{chunk}} = O(n)
\]

\subsection{Stage 2: Per-Segment Analysis (No Batched Worker)}
Each segment $C_i$ is analyzed independently; there is no single batched worker call over all chunks.

\paragraph{Segment-code variant.}
Optionally generate code once:
\[
Q = \mathcal{G}_{\text{code}}(x'_{1:K})
\quad (\text{or } Q = \text{Cache}[\kappa(x')] \text{ if cache hit})
\]
Code $Q$ reads $\text{context}[\text{segment}]$, builds prompt, calls $\text{llmquery}(p)$, parses response, and sets $\text{segmentresult}$.

For $i=1,\ldots,k$: set
\[
\text{ctx}_i = \{\text{segment}: C_i,\ \text{segmentid}: i,\ \text{totalsegments}: k\}
\]
run $\text{REPL}.\text{execute}(Q)$ with $\text{context}=\text{ctx}_i$; read $\text{segmentresult}$ from REPL $\Rightarrow \text{response}_i$.

\paragraph{RLM per-segment variant.}
For $i=1,\ldots,k$:
\[
\text{ctx}_i = \{\text{segment}: C_i,\ \text{segmentid}: i,\ \text{totalsegments}: k\}
\]
\[
\mathit{comp}_i = \text{RLM}.\text{completion}(\text{ctx}_i, \text{rootprompt})
\]
\[
\text{response}_i = \mathit{comp}_i.\text{response}
\]
(string form of $\text{segmentresult}$).

\paragraph{Unified.}
\[
\text{response}_i = \text{SegmentAnalysis}(C_i, i, k)
\]
Each verdict is a tuple:
\[
v_i = (y_i, \gamma_i, A_i, e_i)
\]
where
\[
y_i \in \mathcal{V},\quad
\gamma_i \in [0,100],\quad
A_i \subseteq \mathcal{F},\quad
e_i \in \text{String}.
\]
Time (parallel up to $w$ segment workers):
\[
T_{\text{analysis}} = O\!\left(\frac{k\cdot T_{\text{worker}}}{\min(w,k)}\right)
\]

\subsection{Stage 3: Result Parsing}
Parse each LLM response into a structured verdict:
\[
v_i = \mathcal{P}(\text{response}_i) =
\begin{cases}
\text{JSON}_{\text{parse}}(\text{response}_i) & \text{if valid JSON} \\
\text{literaleval}(\text{response}_i) & \text{if valid dict repr} \\
\text{SafeDefault}() & \text{otherwise}
\end{cases}
\]
\[
\text{SafeDefault}() = (\text{SAFE}, 0, \text{"Segment classified as SAFE."}, \emptyset)
\]
Apply $\text{NormalizeExplanation}(e_i)$ to replace parse-failure phrases.
\[
T_{\text{parse}} = O(k)
\]

\subsection{Stage 4: Verdict Aggregation and Rule-Based Overrides}
Malicious chunk count:
\[
m = \sum_{i=1}^{k} \mathbb{1}_{y_i=\text{MALICIOUS}}
\]

\paragraph{Conservative aggregation (no strategy parameter).}
\[
V^{(0)} =
\begin{cases}
\text{MALICIOUS} & \text{if } m>0 \\
\text{SAFE} & \text{otherwise}
\end{cases}
\]

\paragraph{Confidence (initial).}
\[
\gamma^{(0)} =
\begin{cases}
\max_{i:\,y_i=\text{MALICIOUS}} \gamma_i & \text{if } V^{(0)}=\text{MALICIOUS} \\
0 & \text{otherwise}
\end{cases}
\]

\paragraph{Attack vector union (initial).}
\[
A^{(0)} = \bigcup_{i:\,y_i=\text{MALICIOUS}} A_i
\]

\paragraph{Rule-based overrides (applied to full text $x'$).}
\begin{itemize}
\item \textbf{Indirect weapon signals:}
If $V^{(0)}=\text{SAFE}$ and $\text{IndirectWeaponSignals}(x')$, then
\[
V^{(1)}=\text{MALICIOUS},\quad
\gamma^{(1)}=\max(\gamma^{(0)},85),
\]
\[
A^{(1)} = A^{(0)} \cup \{\text{indirectweaponinstructions},\ \text{contentpolicy}\}
\quad (\text{if } A^{(0)}=\emptyset).
\]

\item \textbf{Reporting context:}
If $V^{(1)}=\text{MALICIOUS}$ and $\text{ReportingContextSignals}(x')$ and $\neg\text{IndirectWeaponSignals}(x')$, then
\[
V^{(2)}=\text{SAFE},\quad
\gamma^{(2)}=\min(\gamma^{(1)},30),\quad
A^{(2)}=\emptyset.
\]

\item \textbf{Base64 escalation:}
If $\mathit{obf}=\text{"base64"}$ and $V^{(\cdot)}=\text{SAFE}$:
\begin{itemize}
\item If $\text{JailbreakKeywordsInBytes}(\text{Base64ToBytes}(x))$ then set
\[
V=\text{MALICIOUS},\quad \gamma=\max(\gamma,85),
\]
and add $\{\text{base64obfuscation},\ \text{contentpolicy}\}$.
\item Else if decoded $x'$ has low readability ($\text{Readability}(x') < \tau_{\text{low}}$) then set
\[
V=\text{MALICIOUS},\quad \gamma=\max(\gamma,75),
\]
and add $\{\text{base64obfuscation},\ \text{suspiciousencoding}\}$.
\end{itemize}
\end{itemize}

\paragraph{Final.}
\[
y_{\text{final}} = V,\quad
\gamma_{\text{final}} = \gamma,\quad
A_{\text{final}} = A
\]
after all overrides.

\subsection{Complete Detection Function}
\[
\boxed{
\begin{aligned}
\Phi: \mathcal{X} &\rightarrow \mathcal{Y} \times [0,100] \times 2^{\mathcal{F}} \times \text{Metadata} \\
\Phi(x) &= (y_{\text{final}}, \gamma_{\text{final}}, A_{\text{final}}, \ldots)
\end{aligned}
}
\]

\subsection{Algorithm decomposition (True RLM flow)}
\[
\begin{aligned}
x' &= \mathcal{D}(x), \quad \mathit{obf} = \text{obfuscation flag}
&& \text{(De-obfuscation)} \\
C_1, \ldots, C_k &= \mathcal{C}(x')
&& \text{(Chunking)} \\
\text{For } i &= 1,\ldots,k \text{ (parallel up to } w \text{):} \\
\quad \text{response}_i &= \text{SegmentAnalysis}(C_i, i, k)
&& \text{(Per-segment analysis)} \\
\quad v_i &= \mathcal{P}(\text{response}_i)
&& \text{(Parsing)} \\
(y_{\text{final}}, \gamma_{\text{final}}, A_{\text{final}})
&= \mathcal{A}(v_1,\ldots,v_k;\ x',\ \mathit{obf})
&& \text{(Aggregation + overrides)}
\end{aligned}
\]
$\mathcal{A}$ here includes conservative merge plus IndirectWeaponSignals, ReportingContextSignals, and base64 escalation as above.

\subsection{Time Complexity}
Total time:
\[
T(x)=T_{\text{deobf}} + T_{\text{chunk}} + T_{\text{code}} + T_{\text{analysis}} + T_{\text{parse}} + T_{\text{agg}}
\]
\[
T_{\text{code}} = 0 \ \text{(RLM per-segment)} \quad \text{or} \quad O(1)\ \text{with cache (segment-code)}
\]
\[
T_{\text{analysis}} = O\!\left(\frac{k\cdot T_{\text{worker}}}{\min(w,k)}\right)
\]
\[
T_{\text{agg}} = O(k)
\]
Asymptotic:
\[
T(x)= O(n) + O(k) + O\!\left(\frac{k\cdot T_{\text{worker}}}{\min(w,k)}\right)
\]
There is no batched worker call; speedup comes from parallelizing the $k$ segment analyses across $w$ workers.

\subsection{Optimization Objective}
Primary (detection rate):
\[
\max_{\theta}\ \mathbb{E}_{(x,y)\sim \mathcal{D}}\left[\mathbb{1}\{\Phi(x)=y\}\right]
\]
Constraints: $\text{FPR} \le \epsilon$, $\mathbb{E}[T(x)] \le T_{\max}$, confidence calibration for MALICIOUS verdicts.

\subsection{Performance Metrics}
\[
\text{Acc} = \frac{\text{TP}+\text{TN}}{\text{TP}+\text{TN}+\text{FP}+\text{FN}}
\]
\[
\text{Precision} = \frac{\text{TP}}{\text{TP}+\text{FP}}
\]
\[
\text{Recall} = \frac{\text{TP}}{\text{TP}+\text{FN}}
\]
\[
\text{F1} = 2\cdot \frac{\text{Precision}\cdot \text{Recall}}{\text{Precision}+\text{Recall}}
\]
\[
\text{FPR} = \frac{\text{FP}}{\text{FP}+\text{TN}}
\]

\subsection{Theoretical Guarantees}
\paragraph{Theorem (Conservative aggregation).}
If any chunk is correctly identified as MALICIOUS, the final verdict is MALICIOUS under conservative aggregation (before overrides).
\[
\exists i:\ y_i=\text{MALICIOUS} \implies V^{(0)}=\text{MALICIOUS}
\]
By definition: $m>0 \implies V^{(0)}=\text{MALICIOUS}$. Rule-based overrides can only escalate SAFE $\to$ MALICIOUS (indirect weapon, base64) or downgrade MALICIOUS $\to$ SAFE (reporting context when no weapon signals).

\subsection{Batch Processing}
For $m$ inputs $x^{(1)},\ldots,x^{(m)}$:
\[
\Phi(x^{(1)}),\ldots,\Phi(x^{(m)})
=
\text{ParallelMap}\bigl(\Phi,\ x^{(1)},\ldots,x^{(m)},\ w_{\text{cross}}\bigr)
\]
Cross-text parallelization:
\[
T_{\text{batch}}(m) \approx \frac{1}{w_{\text{cross}}}\sum_{j=1}^{m} T\!\left(x^{(j)}\right) + O(w_{\text{cross}})
\]

\subsection{Summary: True RLM vs Article (Strategy-Based) Formulation}
\begin{center}
\begin{tabular}{|l|l|l|}
\hline
\textbf{Aspect} & \textbf{Strategy-based (article)} & \textbf{True RLM (this formulation)} \\
\hline
Master output & Strategy $s=(o,c,f,a,r)$ & Code (segment-code) or per-segment RLM (detector\_rlm) \\
\hline
De-obfuscation & $\mathcal{D}(x; s.o)$ & $\mathcal{D}(x)$ detector-side, base64 only \\
\hline
Chunking & $\mathcal{C}(x'; s.c)$ & $\mathcal{C}(x')$ fixed (paragraphs / 1200, single if $|x'|<200$) \\
\hline
Analysis & $\text{LLM}^{\text{batch}}_{\text{worker}}(p_1,\ldots,p_k)$ & Per-segment: $\text{SegmentAnalysis}(C_i,i,k)$ for each $i$ \\
\hline
Aggregation & $\mathcal{A}(\cdot; s.a)$ (conservative / majority / threshold) & Conservative only + rule-based overrides + base64 escalation \\
\hline
$T_{\text{analysis}}$ & $O(1)$ batched (amortized) & $O(k)$ per-segment (parallelized) \\
\hline
\end{tabular}
\end{center}


\begin{thebibliography}{99}
\bibitem{ref1} Patrick Chao, Edoardo Debenedetti, Alexander Robey, Maksym Andriushchenko, Francesco Croce, Vikash Sehwag, Edgar Dobriban, Nicolas Flammarion, George J. Pappas, Florian Tramer, Hamed Hassani, and Eric Wong. Jailbreakbench: An open robustness benchmark for jailbreaking large language models, 2024.
\bibitem{ref2} Sizhe Chen, Julien Piet, Chawin Sitawarin, and David Wagner. Struq: Defending against prompt injection with structured queries, 2024.
\bibitem{ref3} Sizhe Chen, Arman Zharmagambetov, Saeed Mahloujifar, Kamalika Chaudhuri, David Wagner, and Chuan Guo. Secalign: Defending against prompt injection with preference optimization. In Proceedings of the 2025 ACM SIGSAC Conference on Computer and Communications Security, CCS ’25, pages 2833–2847. ACM, 11 2025.
\bibitem{ref4} Edoardo Debenedetti, Ilia Shumailov, Tianqi Fan, Jamie Hayes, Nicholas Carlini, Daniel Fabian, Christoph Kern, Chongyang Shi, Andreas Terzis, and Florian Tram`er. Defeating prompt injections by design, 2025.
\bibitem{ref5} Peng Ding, Jun Kuang, Dan Ma, Xuezhi Cao, Yunsen Xian, Jiajun Chen, and Shujian Huang. A wolf in sheep’s clothing: Generalized nested jailbreak prompts can fool large language models easily, 2024.
\bibitem{ref6} Yangyang Guo, Yangyan Li, and Mohan Kankanhalli. Involuntary jailbreak: On self-prompting attacks, 2025.
\bibitem{ref7} Kuo-Han Hung, Ching-Yun Ko, Ambrish Rawat, I-Hsin Chung, Winston H. Hsu, and Pin-Yu Chen. Attention tracker: Detecting prompt injection attacks in llms, 2025.
\bibitem{ref8} Sahasra Kokkula, Somanathan R, Nandavardhan R, Aashishkumar, and G Divya. Palisade – prompt injection detection framework, 2024.
\bibitem{ref9} Songze Li, Ruishi He, Xiaojun Jia, Jun Wang, and Zhihui Fu. Knowledge-driven multi-turn jailbreaking on large language models, 2026.
\bibitem{ref10} Huawei Lin, Yingjie Lao, Tong Geng, Tan Yu, and Weijie Zhao. Uniguardian: A unified defense for detecting prompt injection, backdoor attacks and adversarial attacks in large language models, 2025.
\bibitem{ref11} Xiaogeng Liu and Chaowei Xiao. Autodan-reasoning: Enhancing strategies exploration based jailbreak attacks with test-time scaling, 2025.
\bibitem{ref12} Xiaogeng Liu, Nan Xu, Muhao Chen, and Chaowei Xiao. Autodan: Generating stealthy jailbreak prompts on aligned large language models, 2024.
\bibitem{ref13} Yi Liu, Gelei Deng, Yuekang Li, Kailong Wang, Zihao Wang, Xiaofeng Wang, Tianwei Zhang, Yepang Liu, Haoyu Wang, Yan Zheng, Leo Yu Zhang, and Yang Liu. Prompt injection attack against llm-integrated applications, 2025.
\bibitem{ref14} Mantas Mazeika, Long Phan, Xuwang Yin, Andy Zou, Zifan Wang, Norman Mu, Elham Sakhaee, Nathaniel Li, Steven Basart, Bo Li, David Forsyth, and Dan Hendrycks. Harmbench: A standardized evaluation framework for automated red teaming and robust refusal, 2024.
\bibitem{ref15} OWASP GenAI Security Project. Llm01:2025 prompt injection, 2025. Accessed 2026-01-18. 11
\bibitem{ref16} Sebastian. Recursive language models: the paradigm of 2026. prime intellect, 2026.
\bibitem{ref17} Xinyue Shen, Zeyuan Chen, Michael Backes, Yun Shen, and Yang Zhang. ”do anything now”: Characterizing and evaluating in-the-wild jailbreak prompts on large language models, 2024.
\bibitem{ref18} Tianneng Shi, Kaijie Zhu, Zhun Wang, Yuqi Jia, Will Cai, Weida Liang, Haonan Wang, Hend Alzahrani, Joshua Lu, Kenji Kawaguchi, Basel Alomair, Xuandong Zhao, William Yang Wang, Neil Gong, Wenbo Guo, and Dawn Song. Promptarmor: Simple yet effective prompt injection defenses, 2025.
\bibitem{ref19} Willis-Owen and David. How to jailbreak gemini 3 in 2025., 2025.
\bibitem{ref20} Qiang Yu, Xinran Cheng, and Chuanyi Liu. Defense against indirect prompt injection via tool result parsing, 2026.
\bibitem{ref21} Qiusi Zhan, Zhixiang Liang, Zifan Ying, and Daniel Kang. Injecagent: Benchmarking indirect prompt injections in tool-integrated large language model agents, 2024.
\bibitem{ref22} A. L. Zhang. Recursive language models. alex l. zhang (blog)., 2025.
\bibitem{ref23} Alex L. Zhang, Tim Kraska, and Omar Khattab. Recursive language models, 2025.
\bibitem{ref24} Andy Zou, Zifan Wang, Nicholas Carlini, Milad Nasr, J. Zico Kolter, and Matt Fredrikson. Universal and transferable adversarial attacks on aligned language models, 2023.

\bibitem{ref25}
Shenyi Zhang, Yuchen Zhai, Keyan Guo, Hongxin Hu, Shengnan Guo, Zheng Fang, Lingchen Zhao, Chao Shen, Cong Wang, and Qian Wang. JBShield: Defending Large Language Models from Jailbreak Attacks through Activated Concept Analysis and Manipulation. In \emph{Proceedings of the 34th USENIX Security Symposium}, 2025. arXiv:2502.07557.

\bibitem{ref26}
Giulio Zizzo, Giandomenico Cornacchia, Kieran Fraser, Muhammad Zaid Hameed, Ambrish Rawat, Beat Buesser, Mark Purcell, Pin-Yu Chen, Prasanna Sattigeri, and Kush R. Varshney. Adversarial Prompt Evaluation: Systematic Benchmarking of Guardrails Against Prompt Input Attacks on LLMs. arXiv:2502.15427, 2025.

\bibitem{ref27}
InjectPrompt, “How to Jailbreak Gemini 3 in 2025.” Accessed: Feb. 12, 2026. [Online]. Available: https://www.injectprompt.com/p/how-to-jailbreak-gemini-3-in-2025


\end{thebibliography}
\end{document}